\documentclass[conference,onecolumn]{IEEEtran}
\usepackage{cite}
\usepackage{amsmath,amssymb,amsfonts}
\usepackage{graphicx}
\usepackage{textcomp}
\def\BibTeX{{\rm B\kern-.05em{\sc i\kern-.025em b}\kern-.08em
    T\kern-.1667em\lower.7ex\hbox{E}\kern-.125emX}}

\usepackage{tabularx}
\usepackage[dvipsnames]{xcolor}
\usepackage{multicol}
\usepackage[linesnumbered,ruled,vlined]{algorithm2e}
\usepackage{listings}
\usepackage{xspace}
\usepackage{todonotes}
\usepackage{footnote}
\usepackage{pgfplots}
\pgfplotsset{compat=1.15} 
\usepackage{url}
\usepackage[font=normalsize,labelfont=bf]{caption}
\usepackage{hyperref}

\usepackage[normalem]{ulem}%For strike through the text
\usepackage{lineno,hyperref}
\modulolinenumbers[5]
\usepackage{float}

\newcommand{\punt}[1]{}
\newcommand{\cmnt}[1]{}

\newtheorem{theorem}{Theorem}

\newtheorem{lemma}[theorem]{Lemma}
\newenvironment{proof}[1][Proofsketch]{\noindent\textbf{#1.} }{} %\rule{0.5em}{0.5em}\\}
\newcommand{\secref}[1]{\textsection\ref{sec:#1}}
\newcommand{\figref}[1]{{Figure}~\ref{fig:#1}}
\newcommand{\listref}[1]{{Listing}~\ref{lst:#1}}
\newcommand{\tabref}[1]{Table~\ref{tab:#1}}

\newcommand{\lref}[1]{Line~\ref{lin:#1}}
\newcommand{\algoref}[1]{{Algorithm~\ref{alg:#1}}}
\newcommand{\subsecref}[1]{\textsection\ref{subsec:#1}}

\newcommand{\remove}[1]{}

\newcommand{\ignore}[1]{}

\newcommand{\op}{operation\xspace}

\newcommand{\conf}{\emph{conflict~node}\xspace}
\newcommand{\rt}{\emph{root}\xspace}
\newcommand{\trash}{\emph{trash}\xspace}
\newcommand{\Trash}{\emph{Trash}\xspace}
\newcommand{\logFile}{\emph{present\_log}\xspace}
\newcommand{\timestamp}{timestamp\xspace}

\newcommand{\clrtxt}[1]{\textcolor{black}{#1}}
\SetCommentSty{mycommfont}

\definecolor{nodeblue}{RGB}{74, 144, 226}
\definecolor{chocolatem}{RGB}{139, 87, 42}
\definecolor{nred}{HTML}{D0021B}
\definecolor{ngreen}{HTML}{417505}
\newcommand\circled[1]{\tikz[baseline=(char.base)]{\node[shape=circle,draw,inner sep=.3pt, color=nodeblue, fill=nodeblue, fill opacity=1] (char) {$\color{white}#1$};}}
\newcommand\circledm[1]{\tikz[baseline=(char.base)]{\node[shape=circle,draw,inner sep=.3pt, color=chocolatem, fill=chocolatem, fill opacity=1] (char) {$\color{white}#1$};}}

\newcommand\circledred[1]{\tikz[baseline=(char.base)]{\node[shape=circle,draw,inner sep=.3pt, color=nred, fill=nred, fill opacity=1] (char) {$\color{white}#1$};}}

\newcommand\circledgreen[1]{\tikz[baseline=(char.base)]{\node[shape=circle,draw,inner sep=.3pt, color=ngreen, fill=ngreen, fill opacity=1] (char) {$\color{white}#1$};}}

\pgfplotsset{
	grid style = {
		dash pattern	= on 0.8mm off .8mm,
		line cap		= round,
		black!45,
		line width		= .5pt
	},
}

\pgfplotsset{
	base/.style={
		title style	= {at={(0.5,-.38)}},
		scale only axis,
		grid		= both,
		width		= 5.8in,
		height		= 3.4in,
		enlarge x limits	= 0.11,
		tick align			= outside,
		axis x line*		= bottom,
		axis y line*		= left,
		minor grid style	= {thin, color = black!5},
		x label style		= {at = {(0.5,-.1)}},
		x tick label style	= {/pgf/number format/1000 sep= },
		y tick label style	= {/pgf/number format/assume math mode},
	}
}
\pgfplotsset{
	nnc/.style={
		ybar		= 6pt,
		bar width	= 25pt, 
		nodes near coords,
	}
}

\definecolor{nblue}{RGB}{46, 134, 193 }
\definecolor{ngr}{RGB}{142, 68, 173}
\definecolor{col1}{HTML}{7D3C98}
\definecolor{col2}{HTML}{bd910f}

\newcommand{\newtext}[1]{\textcolor{black}{#1}}

\newcommand\Sout{\textcolor{red}\bgroup\markoverwith{\textcolor{red}{\rule[.5ex]{4pt}{0.4pt}}}\ULon}

\begin{document}

\title{{An} Efficient Approach to Move Elements in {a} Distributed Geo-Replicated Tree$^{*}$\footnote{$^*$Author sequence follows lexical order of last names. The proposal of this paper has been accepted in CCGrid 2022.}}

\author{\IEEEauthorblockN{Parwat Singh Anjana$^\dagger$, Adithya Rajesh Chandrassery$^\ddagger$, and Sathya Peri$^\dagger$}
	\IEEEauthorblockA{
		$^\dagger$Department of Computer Science and Engineering, Indian Institute of Technology, Hyderabad, India\\
		$^\ddagger$Department of Computer Science and Engineering, National Institute of Technology Karnataka, Surathkal, India\\
		\texttt{cs17resch11004@iith.ac.in, adithyarajesh.191cs203@nitk.edu.in, sathya\_p@cse.iith.ac.in}
	}
}

\maketitle

\thispagestyle{plain}
\pagestyle{plain}

\begin{abstract}
%Replicated tree data structures are extensively used in collaborative applications and distributed file systems where move operation becomes an essential operation frequently performed by the clients. The move operations at different replicas may be safe locally but may not be safe when merged. Hence, it is challenging to implement the move operation when clients perform arbitrary operations concurrently on different replicas, resulting in various bugs. Existing works have shown bugs like data duplication and cycles in the tree.
Replicated tree data structures are extensively used in collaborative applications and distributed file systems, where clients often perform move operations. Local move operations at different replicas may be safe. However, remote move operations may not be safe. When clients perform arbitrary move operations concurrently on different replicas, it could result in various bugs, making this operation challenging to implement. Previous work has revealed bugs such as data duplication and cycling in replicated trees. In this paper, we present an efficient algorithm to perform move operations on the distributed replicated tree while ensuring eventual consistency. The proposed technique is primarily concerned with resolving conflicts efficiently, requires no interaction between replicas, and works well with network partitions. We use the last write win semantics for conflict resolution based on globally unique timestamps of operations. \newtext{The proposed solution requires only one compensation operation to avoid cycles being formed when move operations are applied.} %The proposed solution require only one compensation action to undo the previously moved node, resulting in a cycle. 
The proposed approach achieves an effective speedup of 68.19$\times$ over the state-of-the-art approach in a geo-replicated setting on Microsoft Azure standard instances at three different continents.%The proposed technique is primarily concerned with efficient conflict resolution, requires no cross-replica cooperation, and is highly available during network partitions. We follow the last write win semantics for the conflict resolution based on globally unique timestamps of operations and require one compensation operation to undo the last moved node leading to a cycle. The experimental evaluation shows that the proposed algorithm outperforms the state-of-the-art algorithm in a geo-replicated setting. %Replicated tree data structures are extensively used in collaborative applications and distributed file systems where move operation becomes an essential operation frequently performed by the clients concurrently. The move operations at different replicas may be safe locally but may not be safe when merged. It is challenging to implement move operations when clients perform arbitrary operations concurrently, resulting in various bugs. Previous works have shown bugs like duplication and cycles in the tree. This paper presents an efficient algorithm to perform move operations on the distributed replicated tree while ensuring eventual consistency. The proposed approach does not require cross-replica coordination and hence is highly available even during network partitions. In case of conflicts, we follow the last write wins semantics based on globally unique timestamps to choose the operation to be applied. The proposed protocol requires one compensation operation to undo the last moved node that leads to a cycle. %We provide formal proof to show the convergence of the replicas and the maintenance of the tree structure.
%We compare the performance of the proposed approach against the state-of-the-art in a geo-replicated setting.

%The conflict-free replicated data types (CRDTs) such as tree has become an indispensable component of many distributed applications. The replicated tree structure is commonly used in distributed file systems where clients from various geo-locations work simultaneously. In order to provide high availability, clients should be able to update their replicas concurrently without any coordination. A concurrent move operation (in which a sub-tree can be moved to multiple positions within the tree) is challenging to implement because cycles can arise, and the tree structure can be broken. Previous works have shown bugs in Dropbox (duplication of data) and Google Drive (error due to cycle formation).

\end{abstract}

\begin{IEEEkeywords}
Conflict-free Replicated Data Types, Eventual Consistency, Distributed File Systems, Replicated Tree
\end{IEEEkeywords}
    \section{Introduction} 
\label{sec:intro}
%Modern distributed applications ensure high system up-time and an enhanced user experience by replicating data onto multiple systems. However due to data replication, consistency issues arise in the presence of concurrent reads and writes. The consistency between the replicas must be ensured when the replicated data are mutable. The most commonly used consistency model provides one of the forms of strong, eventual, and causal consistency~\cite{Kulkarni-CausalSpartan-2017}. A mutation occurs instantaneously on all replicas in the strong consistency (strongest consistency) model; however, the replicas may diverge when the network is partitioned. The CAP theorem~\cite{gilbert2002brewer} asserts that a distributed system can only guarantee two of the three properties: consistency (C), availability (A), and partition failure tolerance (P). The operations are employed in a causal order in a causal consistency model that captures causal relationships~\cite{Kulkarni-CausalSpartan-2017}. The replicas may diverge in the eventual consistency model (weakest consistency model); however, they converge to the same state eventually if no new updates are performed at any replica. 

Modern distributed systems ensure high uptime and availability which is commonly achieved by replicating data onto multiple systems. When clients perform concurrent operations, data replication at different replicas may cause consistency issues. \newtext{Various consistency models have been implemented in the literature to ensure the mutability of replicated data. These consistency models are classified into different classes based on the consistency guarantees they provide, such as strong consistency, eventual consistency~\cite{Vogels2008EventuallyConsistent}, and causal consistency~\cite{Kulkarni-CausalSpartan-2017}. The mutation occurs instantly across replicas in the strong consistency model; this is the strongest condition in an ideal setting. However, replicas may diverge when the network is partitioned; consequently, strong consistency is not easy to achieve with network partitions without sacrificing availability. Further, strong consistency suffers from performance overhead due to high synchronization costs when the network is reliable~\cite{abadi2012consistency}.} 

\newtext{%To summarize, strong consistency is the ideal consistency model that can be satisfied for any replicated system; however, it comes with a high-performance cost, and network partition is another essential concern. Therefore, it may not suit replicated systems requiring high availability and scalability with concurrent updates and a guarantee of converging to the same state. 
Strong consistency is the strongest form of consistency for any replicated system; unfortunately, it comes with a considerable performance penalty. As a result, it may not be suitable for replicated systems that require high availability, scalability with concurrent updates and convergence guarantee to a consistent state. As a result, systems designed based on weaker consistency models such as eventual consistency have become popular \cite{Kulkarni-CausalSpartan-2017,Bailis2013EventualConQueue}.} In the eventual consistency model, replicas may diverge for various reasons; however, they eventually converge to the same state if no new updates are performed at any replica~\cite{Werner2009EventualConCA,Bailis2013EventualConQueue,Burckhardt2014EventualConFTPL}.%In this paper, we focus on eventual consistency.

\newtext{%Techniques developed in the literature to handle collaborative application challenges deal with concurrent updates when multiple users simultaneously make changes.
Concurrent updates to various replicas make it very difficult to converge and has been extensively studied in the literature. Numerous approaches have been developed to overcome this problem in several ways. The most prevalent techniques are \emph{operational transformation} (OT)~\cite{ellis1989concurrency,nichols1995high,seifried2012regional} and \emph{conflict-free replicated data types} (CRDTs)~\cite{preguicca2018conflict,shapiro2011comprehensive,shapiro2011conflict}. OT requires a centralized server and an active server connection to modify the replicated file collaboratively.} In contrast, CRDTs do not require a centralized server and allow peer-to-peer editing. CRDTs have become an indispensable component of many modern distributed applications that guarantee some form of eventual consistency~\cite{Kleppmann2020AHM}. Clients update their replicas concurrently without coordination to provide high availability even when the network is partitioned. \newtext{It allows users to operate locally with no lag, even if they are not connected to other replicas. The system eventually becomes consistent when a user synchronizes with other users and devices. %The best part about CRDTs is that they can do everything without a centralized server.
}

%The replicated tree structure is used extensively in distributed file systems. Clients interactively operate on the tree to perform various operations such as renaming, moving, deleting, and adding new files or directories. A node represents directories in the tree CRDT, while a leaf node represents files.

Popular distributed file systems such as Dropbox and Google Drive optimistically replicate data using a replicated tree data model. Clients interactively operate on the tree to perform various operations, such as updating, renaming, moving, deleting, and adding new files or directories. An interior node in the tree represents a directory, while a leaf node represents a file. This distributed file system runs a daemon on the client's machine that keeps track of changes by monitoring the designated directory~\cite{Kleppmann2020AHM,nair2021coordination}. Clients can read and update files locally on their systems, which can then be synchronized with other replicas. Collaborative text editing and graphical editors are examples of distributed systems that often use the replicated tree data model. %The tree data model is often used in distributed applications such as collaborative text and graphical editors. %The clients can read and update the files on their local system, even when their system is offline. Many distributed applications commonly use the tree data model, such as collaborative text and graphical editors. 

The clients can read and update the files offline on their local system, which can later be synchronized with other replicas. Moving nodes is a common operation in such tree-based collaborative applications. In the file system example, the move operation moves files or directories to a new location within the tree. In a collaborative text editor that stores data using an XML or JSON data model, changing a paragraph to bullet points generates a new list and bullet point node. It then moves the paragraph nodes under the bullet point node. Another example is a collaborative graphical editor (Figma~\cite{figmaURL}) where grouping two objects lead to adding a new node in the tree~\cite{Kleppmann2020AHM}. 

%Moving of nodes is a common operation in such, tree-based collaborative applications. In the file system example, the moving of files or directories to new location within the tree can be done using move operation. In the collaborative text editor that uses XML or JSON to store the data, changing a paragraph to bullet points creates a new list and bullet point nodes. It then moves the paragraph nodes under the bullet point node. Another example is a collaborative graphical editor (Figma~\cite{figmaURL}) where grouping two objects lead to adding a new node in the tree~\cite{Kleppmann2020AHM}. 

%A tree structure renders rich text or vector graphics in collaboratively editing as paragraphs, lists, figures, sections, and drawings, typically allowing users to edit information interactively, resulting in an update to the underlying XML or JSON data~\cite{Kleppmann2020AHM}. \spnote{This line is not clear}
%The clients' primary operation is adding, removing, and moving a node within the same tree. 

%Considering the usefulness, we concentrate on the move operation in replicated tree CRDT in this document. A sub-tree is moved to a different location within the tree in the move operation. This operation is challenging to implement because concurrent operations by different clients may lead to cycles; moreover, the tree structure can be broken sometimes~\cite{Bjorner2007,Kleppmann2020AHM,nair2021coordination}. Due to concurrent operations, a concurrency control mechanism is needed to ensure the correctness of the data structure.

This paper focuses on the move operation in the replicated tree CRDT due to its usefulness. The \emph{move operation} moves a sub-tree within the tree. This operation is difficult to implement because concurrent operations by multiple clients may result in cycles; additionally, the tree structure may be broken~\cite{Bjorner2007,Kleppmann2020AHM,nair2021coordination}. Due to the concurrent operations, a concurrency control mechanism is required to ensure the data structure's correctness. Further, ensuring correctness while providing low latency, high throughput, and maintaining high availability can be very challenging. 

\begin{figure}[!t]%[18]{r}{0.41\textwidth}
	\centering
	%\fbox
	{\includegraphics[width=.72\textwidth]{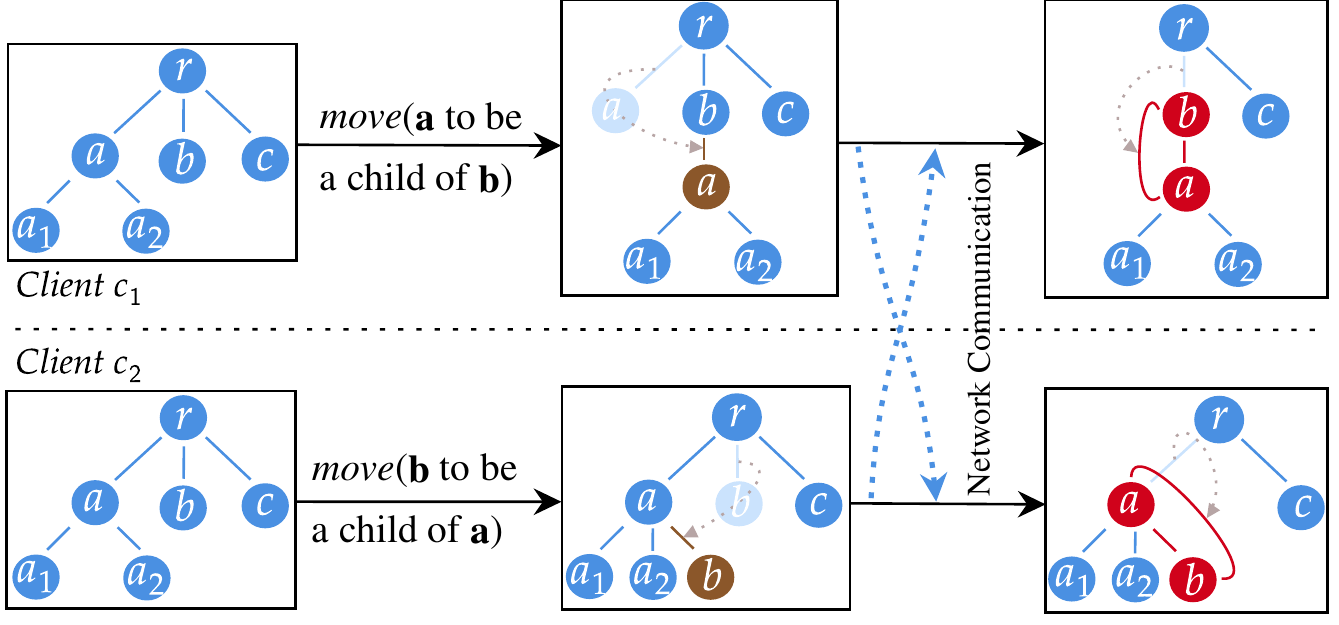}}%\vspace{-.2cm}
	\caption{Difficulty with the move operation on a replicated tree: let us assume a tree structure $t$ rooted at $r$, and two clients $c_1$ and $c_2$, concurrently operating on $t$ in their local replicas. Let say $c_1$ move \emph{($a$ to be a child of $b$)} concurrently with client $c_2$ moving \emph{($b$ to be a child of $a$)} in their local version of $t$ without any coordination. Later, when these replicas are synchronized by propagating local operations, it may produce a cycle $a \leftrightarrow b$ disconnected from the root $r$.}
	\label{fig:cycle}
\end{figure}

An example in \figref{cycle} shows the difficulty associated with the move operation in the replicated setting. The tree structure is replicated on multiple systems. Different clients can perform concurrent operations, leading to various malformations in the tree, such as a cycle, duplication, and detachment from the parent node. Concurrent move operations on the same tree node cause the data duplication problem. Providing support for a concurrent move operation for the replicated tree that does not require continuous synchronization or centralized coordination is problematic because two operations that are individually safe at their local replicas, when combined, might produce a cycle. Prior works by Nair et al.~\cite{nair2021coordination} and Kleppmann et al.~\cite{Kleppmann2020AHM} have shown that Dropbox suffers from duplication, and Google Drive results in errors due to the formation of cycles.

We present an efficient protocol to perform move operations while maintaining the distributed replicated tree structure and ensuring that replicas are eventually consistent. The proposed approach does not require cross-replica coordination and hence is highly available even when the network is partitioned. In case of conflicting operations, we follow the last write win approach based on timestamps computed using the Lamport clock~\cite{Lamport1978clock}, however hybrid logical clock (HLC)~\cite{kulkarni2014HLC} can also be used. The proposed protocol requires one compensation operation to undo the last moved node that causes the cycle. \newtext{Essentially, we address the conflict resolution problem of the replicated tree by minimizing the number of undo and redo operations required to resolve conflicts.}\vspace{.1cm}\\The significant {contributions} are as follows: %\vspace{-.35cm}
\begin{itemize}
    \item %We propose a novel coordination-free, computationally efficient, and low latency move operation on the replicated tree. .The proposed approach supports optimistic replication that allows replicas to temporarily diverge when they are updated but always converge to a consistent state (see \secref{proposed}).
    \newtext{We propose a novel move operation approach on the replicated tree that is coordination-free, computationally efficient, and offers low latency operation. The proposed approach supports optimistic replication, which allows replicas to temporarily diverge during updates but always converges to a consistent state in the absence of new updates (see \secref{proposed}).}%The proposed algorithm supports optimistic replication, allowing replicas to temporarily diverge as they are updated and ensure that they always converge towards a consistent state (see \secref{proposed}).
    \item The proposed algorithm guarantees eventual consistency; correctness proofs are provided in \secref{pc} %Appendix \secref{pc} (due to lack of space) 
    to show the \emph{convergence} of the replicas and the maintenance of the tree structure. 
    \item The performance of the proposed approach is compared against the Kleppmann et al.~\cite{Kleppmann2020AHM}. %state-of-the-art approach~\cite{Kleppmann2020AHM}. 
    \newtext{The experiment results show that the proposed approach achieves an effective speedup of 68.19$\times$ over Kleppmann's approach for the remote move operation (see \secref{results}).}%The experiment results show that the proposed approach outperforms Kleppmann's approach for the remote move operation by 68.19$times$.
    %The experiment results show that the proposed approach outperforms the existing approach~\cite{Kleppmann2020AHM} (see \secref{results}).
\end{itemize}%\vspace{-.15cm}
A brief overview of the related work aligned with the proposed approach is discussed in \secref{rw}, while the system model is given in \secref{sysmod}. \secref{conc} conclude with some future research directions.

    \section{Related Work}
\label{sec:rw}
This section briefly discusses the related work that has been done in line with the proposed approach.

Algorithms on replicated data structures are classified into two classes: operational transformation (OT)~\cite{ellis1989concurrency,jungnickel2016simultaneous,nichols1995high,seifried2012regional} and conflict-free replicated data types (CRDTs)~\cite{preguicca2018conflict,shapiro2011comprehensive,shapiro2011conflict}. Many of the proposed approaches mainly support two operations: insert and delete. Furthermore, the approaches based on OT require a central server and an active network connection between the client and the central server. It means it does not work when the replica is offline. In comparison, CRDT mitigates this issue by allowing asynchronous peer-to-peer communication between replicas using optimistic replication. In the presence of faults to reduce operation response time and increase availability, the optimistic replication~\cite{Saito2005Optimistic} allows replicas to diverge temporarily.  

The CRDTs are mainly categorized into two types: state-based~\cite{shapiro2011comprehensive} and operation-based~\cite{baquero2014making,shapiro2011comprehensive}. The former, alternatively referred to as convergent replicated data types, is more straightforward to design and implement. However, a significant disadvantage is that it requires the transmission of the entire state to every other replica. On the other hand, later, also referred to as commutative replicated data types, transmit only update operations to each replica, thus requiring less network bandwidth. Nonetheless, it is built on the assumption of a reliable communication network, which means that no operations are dropped or duplicated. %The proposed solution is an example of operation-based CRDT. 
The data structures supported by these CRDTs are counters, lists, registers, sets, graphs, trees, etc. The proposed protocol implements efficient move operation on operation-based tree CRDT.

Extensive research has been done to implement geo-replicated distributed tree data structures, which are used in a variety of distributed applications, including Google Drive, Dropbox (file systems), Google Docs, Apple's Notes App (collaborative text editing), and Figma (collaborative graphical editor). For replicated tree CRDTs, numerous algorithms have been proposed. Several of these algorithms support only insert and delete operations and have a long response time or latency. 

Martin et al.~\cite{martin2010scalable} proposed tree CRDT for XML data to support insert and delete operation. While supporting these two operations in JSON data format is proposed by Kleppmann et al.~\cite{kleppmann2017conflict}. Insert and delete operations can be used to implement a move operation; however, this could lead to the data duplication problem and increase the number of computation steps. A replicated file system using tree CRDT is implemented in~\cite{najafzadeh2016analysis,Najafzadeh2018CoDesign}. These solutions result in data duplication (tree node duplication). When replicas perform operations concurrently, data duplication occurs, resulting in irreversible divergence between replicas or the need for manual intervention to restore replicas to a consistent state.

Data duplication is a severe issue, and for large systems, manually handling this problem is very difficult. %These solutions lead to data (tree node) duplication. When replicas execute operations simultaneously, data duplication occurs, resulting in irreversible divergence between replicas or requiring manual effort to bring replicas back to a consistent state. Data duplication is a severe issue, and large systems cannot be handled manually. %Data duplication occurs when replicas execute operations concurrently, resulting in permanent divergence between replicas or requiring manual effort to make replicas back to the consistent state. Data duplication is a serious problem, and manual effort is not feasible for large systems.
%For replicated tree CRDTs, numerous algorithms have been proposed. Several of these algorithms are based on locking schemes. The lock-based system suffers from high operation latency or response time while other approaches supports only insert and delete operation. Alternatively, when more than one replica performs the move operation on the same node, data duplication occurs, resulting in permanent divergence between the replicas or requiring manual effort to get the replicas back to the same state. Data duplication is a serious problem, and manual exercises are not recommended.%There have been several solutions in the literature that proposed the solution for replicated tree CRDTs. However, several tree replication approaches are based on locking techniques. The lock-based solution increases the latency or operation response time for the move operation. Alternatively, suffer from data duplication when more than one replica operates on the same node to move, resulting in permanent divergence between the replicas or require manual effort to transform the replicas to the same state. Data duplication is a serious issue, and manual exercises are not desirable.
\newtext{On the other hand, some techniques require extensive metadata exchanges between replicas to mitigate these issues, which increases network bandwidth requirements. The solution proposed in~\cite{Tao2015merging} results in a directed acyclic graph on concurrent moves. The approach proposed in~\cite{nair2021coordination} requires causal delivery of operations that may not be possible when a replica crash fails, or the network is unreliable.}

\newtext{As discussed earlier in \secref{intro}, move operations are difficult to implement because two concurrent moves can produce a cycle and separate the node from its parent or ancestor. Moving a node in its descendent tree may produce a cycle and break the tree structure. Local move operations on a replica may or may not result in cycles. However, remote move operations may cause cycles that must be handled properly to preserve consistency. Hence, an efficient approach must be proposed to move the nodes and their subtrees to another location in the replicated tree. %As discussed earlier in \secref{intro}, the move operations are challenging to implement because two concurrent moves may lead to a cycle and disconnect the node from their parent / ancestor. Moving a node to a different location in its descendant tree is not supported because it may end up causing a cycle and breaking the tree structure. Local move operations on a replica may or may not result in cycles. However, remote move operations could result in cycles that must be appropriately handled to maintain consistency across the replicas. Hence, an efficient approach must be proposed to move the nodes and their subtrees to another location in the replicated tree.
}
%Additionally, some approaches require high metadata exchanges between the replicas, increasing network bandwidth requirements. Moreover, the solution proposed in~\cite{Tao2015merging} results in the directed acyclic graph on concurrent moves. While the approach proposed in~\cite{nair2021coordination} requires causal delivery of operations, that may not be possible when a replica crash fails, or the network is unreliable. The approach that improves the efficiency of the move operation is one of the significant challenges because two concurrent moves may lead to a cycle and detach the node from their parent / ancestor, as discussed earlier in \secref{intro}. Moving a node to the location in its decent tree is not allowed because it may lead to the cycle and break the tree structure. Moreover, an efficient approach is required to move the node with its sub-tree to another location in the tree. 

%The most recent work in~\cite{Kleppmann2020AHM} is computation-intensive and requires a lot of compensation operation to avoid the cycles in the tree. Their approach relies on making a total global ordering of operations to ensure consistency. Before applying any operation, they first undo all operations applied, which is having a higher timestamp than the received operation, and then apply the received operation. Further, redo all those undone operations. They maintain total order because some operations cannot be applied as they will result in a cycle. So by having the same final order, the operations ignored will be the same, and all replicas converge to the same final state.

The most recent work is proposed by Kleppmann et al.~\cite{Kleppmann2020AHM}. This approach is computation-intensive, requires many compensation operations to avoid the cycles in the replicated tree, and relies on making a total global ordering of operations to ensure strong eventual consistency. In Kleppmann's approach, before applying any remote operation (operation received from remote replica), they first undo all operations applied with a higher timestamp than the received operation, then apply the received operation, finally, redo all those undone operations. They maintain total global order between the operations and ensure strong eventual consistency. %Because some operations cannot be applied, resulting in a cycle. So by having the same final order, the operations ignored will be the same, and all replicas eventually converge to the same final state. 
\newtext{Unlike their approach, which requires multiple undo and redo operations per remote move operation, the proposed approach requires only one undo and compensation operation per conflicting operation and avoids multiple undo and redo operations for non-conflicting move operations.}

\newtext{The proposed approach ensures eventual consistency. It avoids re-computation for non-conflicting changes to the tree by identifying which changes might cause problems to arise. In our approach for a remote operation that creates a problem (cycle), we undo one operation and send that as a compensation operation to all other replicas. By doing this, we save the time of re-computation for non-conflicting operations, as well many operations that need to be undone and redone can be avoided. Essentially, the number of compensation operations is just 1 per cycle and 0 for safe operations. We observed that such a simple approach improves the performances significantly. Additionally, }there is no data duplication in the proposed approach and does not result in directed acyclic graph on concurrent move operation that may lead to inconsistency or divergence between the replicas. The additional metadata, causal delivery, and strict global total order while applying the operations for consistency are not required. So, we propose a novel coordination-free efficient move operation on replicated tree CRDT to support low latency and high availability operations.

    \section{System Model}
\label{sec:sysmod}
%This section presents the system model of the proposed approach. 

Following system model in~\cite{Kleppmann2020AHM}, there are $n$ \emph{replicas} ($r_1, . . . , r_n$) communicate with each other in a completely asynchronous in a peer-to-peer fashion. %There is no central server or consensus protocol. \spnote{why is consensus protocol important here?} 
We assume that replicas can go offline, crash, or fail unexpectedly. Each replica is associated with a \emph{client}. Each client performs operations on their local replicas. Each operation is then communicated to all other replicas asynchronously via messages. A message may suffer an arbitrary network delay or be delivered out of order. Clients can read and update data on their local replica even when the network is partitioned or their replica is offline. 

We consider a replicated tree structure $t$ rooted at \rt{} to which clients add new nodes, delete nodes and move the nodes to the new location within the tree. %\Sout{An internal node of the tree represents directories in the tree, while a leaf node represents files.}
\clrtxt{In a file system, an internal node of the tree represents directories, while a leaf node represents files. In collaborative text editing, different sections, paragraphs, sentences, words, etc., in the document can be represented as tree nodes. } %\spnote{Is this assumption always correct: a node represents directories while leaf represents files?}

%We propose an efficient algorithm to maintain the replicated tree structure. The proposed algorithm is executed on each replica $r_i$ without any distributed shared memory to operate on tree $t$. Clients generate the operations and apply them on their local replica; and, every operation is communicated asynchronously via the network to all other replicas. On receiving any operation, remote replicas apply operations using the same algorithm. The proposed algorithm supports three operations on the tree:%\vspace{-.2cm}
We propose an efficient algorithm to maintain the replicated tree structure. The proposed algorithm is executed on each replica $r_i$ without any distributed shared memory to operate on tree $t$. Clients generate the operations, apply them on their local replica, and communicate them asynchronously via the network to all other replicas. On receiving an operation, remote replicas apply them using the same algorithm. The proposed algorithm supports three operations on the tree:
\begin{itemize}
	\item \emph{Inserting} a new node in the tree.
	\item \emph{Deleting} a node from the tree.
	\item \emph{Moving} a node along with a sub-tree to a child of a new parent in the tree.
\end{itemize}

All three operations can be implemented as a move operation. The \emph{move} operation is a tuple consisting of \timestamp{} `ts', node `n', and new parent `p', i.e., \emph{move$\langle$ts, n, p$\rangle$}. The \timestamp{} `ts' is unique and generated using Lamport \timestamp{s}~\cite{Lamport1978clock}, the node `n' is the tree node being moved, and the parent `p' is the location of the tree node to which it will be moved. We represent node \timestamp{} as `node.ts', and operation \timestamp{} as `o.ts'.

For example, $move_x\langle ts_i,\ n_j,\ p_k \rangle$ means that a node with id $n_j$ is moved as a child of a parent $p_k$ in the tree $t$ at time $ts_i$ in replica $r_x$. The additional information about the old parents of the node being moved is also logged in the \logFile{} used to undo the cyclic operations when cycles are formed due to the move of the $n_j$ by the clients at different replicas. The move operation removes $n_j$ from the current parent and moves it under the new parent $p_k$ along with the sub-tree of $n_j$; however, if $n_j$ does not exist in the tree, a new node with $n_j$ is created as a child of $p_k$. 

We implement \emph{insert} and \emph{delete} operations as a \emph{move} operation. \emph{Insert} is implemented as a move operation, where the node being moved ($n_j$ in the above example) does not already exist in the tree. 
For \emph{delete}, we use a special node called \trash{}\clrtxt{$-$a child of root and }the parent of all deleted nodes. When a delete on a node $n_j$ is invoked, it is moved to the sub-tree of \trash. We explain further details in \secref{proposed}.
% moving to where the node will be created with a unique id. 

The proposed algorithm satisfies the consistency property \emph{convergence}$-$two replicas are said to converge or be in the same state if both of them have seen and applied the same set of operations. The replicas may apply the operations in any order due to reordering of messages and delays. This implies operations must be commutative. The formal proof is provided in \secref{pc}.
    \section{Proposed Algorithm}
\label{sec:proposed}
This section describes the proposed algorithm for performing efficient move operations on the replicated tree. Each replica is modeled as a state machine that transitions from one state to the next by performing an operation. There is no shared memory between replicas, and the algorithm operates autonomously. The proposed protocol requires no central server or consensus mechanism for replica coordination, requires minimal metadata, and satisfies \emph{eventual consistency}. A client generates and applies operations locally with \algoref{applylocal}, then sends them asynchronously over the network to all other replicas with \algoref{Propagate}. We have described the main idea while all the details are explained as pseudocode in the various algorithms.

The proposed algorithm supports \emph{insert, delete} and \emph{move} operations on a replicated tree. It can be shown that inserting and deleting involve changing various nodes' parents. \emph{insert} can be viewed as the creation of a new node that is to be moved to be the child of a specified parent. For a \emph{delete} operation, the node is moved to be a child of a special node denoted as the \trash{}. Thus, the move operation can be used to implement the other two operations. Hence, in this discussion, we only consider an efficient way of moving nodes.

%Coming to the implementation of other operations using \move operation, 

Each move operation takes as arguments: the node to move and the new parent. Further, each tree node also maintains the \timestamp{} (ts) of the last operation applied which is passed to the move operation.  A move operation is formally defined as: \emph{move$\langle$ts, n, p$\rangle$}. Here $ts$ is the \timestamp{} of the move operation, $n$ is the node to be moved and made as a child of $p$. The data structures are shown in \listref{ds}.

{\algoref{init} is used to initialize the system.} A user generates \op{s} and applies them using \algoref{applylocal} locally, and then sends them asynchronously over the network to all other replicas using \algoref{Propagate}. An operation (local or remote) is applied using \algoref{applylocal} and \algoref{apply}. \algoref{apply} first compares the operation \timestamp{} (i.e., $o.ts$) with the \timestamp{} of the node to be moved (i.e., $node.ts$). It applies the operation only if the $o.ts$ is greater than the $node.ts$. We prove convergence by showing that all the tree nodes across the replicas will get attached to the parent with the latest $ts$. \newtext{Next, we explain how the cycle is prevented in the proposed approach.}

{%\footnotesize
%backgroundcolor=\color{white}, 
\begin{lstlisting}[caption={Data Structures}\label{lst:ds}]
move {
  clock ts; //Unique timestamp using Lamport clock.
  int n; //Tree node being moved.
  int p; //New parent.
};
treeNode {
  int id; // Unique id of the tree node.
  clock ts; //Timestamp of the last operation applied on the node.
  int parent; //Parent node id.
};
lc_time - Lamport timestamp of a replica.
root - Original starting point of the tree.
present_log - Stores m unique previous parents of each node in the adjacency list form.
ts - Timestamp.
ch[] - Array of channels (size equal to the number of replicas).
\end{lstlisting}
}
%\end{minipage}
\setlength{\intextsep}{0pt}
\begin{algorithm}[!ht]
	\caption{$init()$: Initialize the system and create threads.}
	\label{alg:init}
	%\SetAlgoLined
	\setcounter{AlgoLine}{0}
	\SetKwFunction{init}{$init$}
	\SetKwProg{Pn}{Procedure}{:}{ }
	\Pn{\init { }}{
		\tcp{Create number of thread equals to the number of replicas, these thread call send function and in loop.}
		
		lc\_time $\leftarrow$ 0; \tcp{For time stamping}
		
		\logFile{} $\leftarrow$ \{\}; \tcp{Adjacency list to store previous parents.}
		
		\For{i = 0 to numReplicas} {
			\tcp{Create `numReplicas' threads to send local \op{s} to remote replicas.}
			
			thread(send, ch, i); %\tcp{$ch[i]$ is an \op{} thread $i$ will send to replica r$_i$.}
		}
		\tcp{Create one thread to generate and apply local \op{s}.}
		thread(localGenerator);
		
		\tcp{Create `numReplicas' receiver threads to receive and apply remote \op{s}.}
		\For{i = 0 to numReplicas}{
			thread(applyRemote, j);
		}
	}
\end{algorithm}
\begin{algorithm}[!ht]
%\footnotesize
\caption{\texttt{send}(channel ch, i): send local operations to other replicas.}
\label{alg:Propagate}
\setcounter{AlgoLine}{15}
\SetKwProg{Pn}{Procedure}{:}{}
\SetKwFunction{send}{\texttt{send}}
\Pn{\send{channel ch, i}}{
	\While{true} {
		
		move$\langle$ts, n, p$\rangle$ $\leftarrow$ ch.get(i);
		
		\If{ts == -1} {
			break; \tcp{thread will join when this condition becomes true.}
		}
		RPC.send(move$\langle$ts, n, p$\rangle$);
	}
}
\end{algorithm} 
\begin{algorithm}[!ht]
\caption{\texttt{checkCycle}($n_i$, $n_j$): detect cycle between two nodes in the tree `t'.}
\label{alg:checkCycle}
%\vspace{.1cm}
%\footnotesize
%\SetAlgoLined
\setcounter{AlgoLine}{21}
%\KwResult{ }
\SetKwProg{Pn}{Procedure}{:}{}
\SetKwFunction{checkCycle}{\texttt{checkCycle}}
\Pn{\checkCycle{$n_i$, $n_j$}}{
	\While{$n_j$ $\neq$ \rt{}}{
		\If{$n_j$ == $n_i$}{
			\Return{TRUE}; 
		}
		$n_j$ $\leftarrow$ get\_node(\rt{}, $n_j.parent$);
	}
	\Return{FALSE};
}
\end{algorithm}
\begin{algorithm}[H]
%\footnotesize
\caption{\texttt{findLast}($n_i, n_j$): find node with highest \timestamp{} in cycle.}
\label{alg:findLast}
\setcounter{AlgoLine}{27}
\SetKwProg{Pn}{Procedure}{:}{}
\SetKwFunction{findLast}{\texttt{findLast}}
\Pn{\findLast{$n_i$, $n_j$}}{
	
	maxTS $\leftarrow$ $n_i.ts$;

	undoNode $\leftarrow$ $n_i$;

	\While{$n_j$ $\neq$ $n_i$}{
	
		\If{$n_j.ts$ $>$ maxTS} {

			undoNode $\leftarrow$ $n_j$;

			maxTS $\leftarrow$ $n_j.ts$;
		}
	
		$n_j$ $\leftarrow$ get\_node(\rt{}, $n_j.parent$);
	}
	\Return undoNode;
}
\end{algorithm}

\begin{algorithm*}%[!ht]
%\footnotesize
\caption{\texttt{applyLocal}($n$, $p$): apply local operations and send them to other replicas.}
\label{alg:applylocal} 
\setcounter{AlgoLine}{36}
\SetKwProg{Pn}{Procedure}{:}{}
\SetKwFunction{local}{\texttt{applyLocal}}
\Pn{\local{$n, p$}}{
	%timer.start();

    $node_n$ $\leftarrow$ get\_node(\rt{}, n); \tcp{Gets reference of the node with id $n$ in $t$.} 
    
    $node_p$ $\leftarrow$ get\_node(\rt{}, p); \tcp{Gets reference of the node with id $p$ in $t$.} 

	Lock();\label{lin:algo1-lock}\tcp{Get lock, so that at a time only one operation will be applied by threads (local thread or receiver thread) on the tree `t'.}
	
	ts $\leftarrow$ ++lc\_time;\label{lin:algolc-lc-time}
	
	\tcp{$checkCycle()$: checks for the cycle and returns $true$ if cycle found.}

	\If{!checkCycle($node_n$, $node_p$) }{\label{lin:algolc-cc}
	    \logFile{}[$node_n$.id].add($node_n$.parent); \tcp{Update the current parent of $node_n$ in the \logFile{}.}\label{lin:algolc-updatelog}
	    
		$node_n$.parent $\leftarrow$ $node_p$.id; 
		
		$node_n$.ts $\leftarrow$ ts;
	}\label{lin:algolc-apply}
	Unlock();\label{lin:algo1-unlock}
	
	%timer.end();
	
	\tcp{Send move operation to other replicas.}
	
	\For{j = 0 to numReplicas}{
		ch[j].add(move$\langle$ts, n, p$\rangle$); 
	}
}

\end{algorithm*}
\begin{algorithm*}%[!ht]
%\footnotesize
\caption{\texttt{applyRemote}(): receives and applies remote move operations.}
\label{alg:apply}
\setcounter{AlgoLine}{48}
\SetKwProg{Pn}{Procedure}{:}{}
\SetKwFunction{update}{\texttt{applyRemote}}
\Pn{\update{ }}{
	\While{true}{
		move$\langle$ts, n, p$\rangle$ $\leftarrow$ Stream.Receive();  \tcp{Receive remote move operation. If it returns $End$, receiver threads will stop.}
		\If{End} {
			break;
		}
		
		%timer.start();
		
        $node_n$ $\leftarrow$ get\_node(\rt{}, n); \tcp{Gives reference of the node with id $n$ in $t$.} 
    
        $node_p$ $\leftarrow$ get\_node(\rt{}, p); \tcp{Gives reference of the node with id $p$ in $t$.} 
		
		Lock();\label{lin:algo5-lock} \tcp{Get lock, so that at a time only one operation will be applied by threads (local thread or receiver thread) on the tree `t'.}
		
		\tcp{Already applied an operation with higher timestamp on n node then ignore the received operation with smaller timestamp since node timestamp will be higher.}
		\If{ts $<$ $node_n$.ts}{\label{lin:algo1-16}
			\Return;\label{lin:algo1-17}
		}
		
		lc\_time $\leftarrow$ max(ts, lc\_time);\label{lin:algo1-18-2}

        \logFile{}[$node_n$.id].add($node_n$.parent); \tcp{Update the current parent of $node_n$ in the \logFile{}.}\label{lin:algo1-18-1}

		\tcp{$checkCycle()$: checks for the cycle and returns $true$ if cycle found.}
		
		\If{checkCycle($node_n$, $node_p$)}{\label{lin:algo1-19} 
			\tcp{Find the node between $node_n$ - $node_p$ with highest timestamp.}
			
			$undoNode$ $\leftarrow$ findLast($node_n$, $node_p$);\label{lin:algo1-20}
			
			\tcp{Undo (move back) to a previous parent, not in the sub-tree of $node_n$. Keep searching till it gets a suitable node to undo; if not found safe previous parent, then move undoNode under \conf{}.}
			
			\While{true} {\label{lin:algo1-wloop}
				
				\If {\emph{\logFile{}}[undoNode.id] == NULL}{
					
					$undoParent$ $\leftarrow$ \conf{};\label{lin:algo1-20-2}
				}
				\Else{
					\tcp{Get and delete the previous parent from the \logFile{} for undoNode.}
					
					$undoParent$ $\leftarrow$ \logFile{}[$undoNode$.id].pop();
				}
				\tcp{Check if a cycle exists between $n$ and $undoParent$; if no cycle, then found a safe node to move back that breaks the cycle between $n$ and $p$.}
				
				\If{!checkCycle(node$_n$, undoParent) } {
					break;
				}
			}\label{lin:algo1-wloopend}

			applyLocal($undoNode$, $undoParent$);\label{lin:algo1-21-2}
			
			\tcp{Send the compensation operation to other replicas}
			\For{j = 0 to numReplicas}{
				ch[j].add(move$\langle$ts, n, p$\rangle$); \label{lin:algo1-23}
			}
		}
		
		\tcp{Already applied a higher timestamp operation.}
		
		\If{undoNode == $node_n$}{ \label{lin:algo1-24}
			
			Unlock();
			
			%timer.end();
			
			\Return;\label{lin:algo1-25}
		}
		
		%\tcp{Apply received remote move operation.}
		
		$node_n$.parent $\leftarrow$ $node_p$.id;\label{lin:algo1-final1}

		$node_n$.ts $\leftarrow$ ts; \label{lin:algo1-26}

		Unlock();\label{lin:algo5-unlock}
		
		%timer.end();
		
	}
}
\end{algorithm*}

%\Sout{From \figref{cycle}, let us assume that initially, both clients (replicas) consist of the same tree with \timestamp{} as shown in \figref{proposed} (sub-figure (i) and (ii)).}
%\clrtxt{ From \figref{cycle}, let us assume that initially, both clients (replicas) consist of the same tree with \timestamp{} as shown in \figref{proposed} (sub-figure (i) and (ii)).}

%\begin{minipage}{\linewidth}

%It is easy to prove convergence since all the tree nodes across replicas will be attached to the parent with the latest ts. A new node is created for the \emph{insert} operation that is moved to the child of the specified parent. For a \emph{delete} operation, the node is moved to be a child of \trash{}.

%The operations are of the form \emph{move$\langle$ts, n, p$\rangle$}. %Where $t$ is a globally unique \timestamp{} for the operation, $n$ is the tree node to be moved, and $p$ is the tree node that will be the parent of $n$.

\vspace{.15cm}
\noindent
\textbf{Preventing Cycles: }Recall from \secref{intro} that a cycle is formed when an ancestor tree node becomes a child of its descendant tree node. Preventing cycles is difficult because concurrent move operations on different replicas may be safe independently. However, a cycle may be formed when move operations from different replicas are merged. To avoid cycles, the proposed algorithm uses \timestamp{s} and compensation operations. We check for cycles prior to performing any operation by determining whether the node to be moved is an ancestor of the new parent. We check for each operation to avoid the formation of a cycle and maintain the tree structure during concurrent moves. Another check identifies the node with the latest \timestamp{} when a cycle is detected (using \algoref{checkCycle}). As a result, the node with the most recent \timestamp{} is returned to its previous parent, which is safe. A previous parent is said to be safe if it is not in the sub-tree of the node to be moved in the move operation.

%Recall that we had discussed cycle formation in \secref{intro}; a cycle is formed when an ancestor tree node becomes a child of its descendant tree node. It is challenging to prevent cycles because concurrent move operations at different replicas may be safe individually. However, when the updates are merged, they can result in a cycle~\cite{Kleppmann2020AHM}. The proposed algorithm uses \timestamp{} and compensation operations to prevent cycles. Before applying any operation, we first check for cycles by seeing if the node to be moved is an ancestor of the new parent. We essentially check for each operation to restrict forming a cycle and maintain the tree structure in concurrent moves. Another check is done to identify the node with the latest \timestamp{} when a cycle is detected (using \algoref{checkCycle}). Accordingly, the node with the latest \timestamp{} is moved back to its previous parent, where it is safe. A previous parent is said to be safe if it is not in the sub-tree of the node to be moved in the move operation. 

\algoref{applylocal} and \algoref{apply} ensure that all operations applied will not form a cycle. \algoref{applylocal} applies the local move operations, while \algoref{apply} applies remote move operations.\footnote{\newtext{We implemented separate processes for performing local and remote move operations on a replica. We employ the \texttt{Lock}() and \texttt{Unlock}() methods in \algoref{applylocal} at \lref{algo1-lock}, \ref{lin:algo1-unlock}, and at \lref{algo5-lock}, \ref{lin:algo5-unlock} in \algoref{apply} to synchronize these processes.}} The procedure followed to apply a remote move operation by \algoref{apply} is explained here. Before applying a move operation (i.e., \emph{move$\langle$ts, n, p$\rangle$})  \algoref{apply} checks if the operation{'s} \timestamp is greater than the previous move operation's \timestamp{} applied on $node_n$ at \lref{algo1-16}. If the  \timestamp of the operation to be applied is smaller, it ignores the operation at \lref{algo1-17}. The following steps occur when a cycle is detected. The algorithm finds the node with the highest \timestamp in the cycle and assigns it to \emph{undoNode} at \lref{algo1-20}. Then find a safe previous parent for the \emph{undoNode} in the while loop from \lref{algo1-wloop}. If there are no more previous parents left, then keep the previous parent as a \conf{} at \lref{algo1-20-2}. Then move the \emph{undoNode} to be a child of safe previous parent \emph{undoParent} using \algoref{applylocal} at \lref{algo1-21-2}. This internally updates the Lamport \timestamp{}, which will be used for the undo operation at \lref{algolc-lc-time}. Then the operation is applied in the local replica at \lref{algolc-cc} and \lref{algolc-apply}. After that, it send the compensation operation to other replicas at \lref{algo1-23}. If \emph{undoNode} is the same as $node_n$ at \lref{algo1-24} then return at \lref{algo1-25}. Since it already applied an operation with a higher \timestamp{} on $node_n$ as the undo operation. Otherwise, apply the operation to change the parent of $node_n$ at Lines \ref{lin:algo1-final1} and \ref{lin:algo1-26}.

\begin{figure}[!t]%{b}{0.4\textwidth}
	\centering
	%\fbox
	{\includegraphics[width=.97\textwidth]{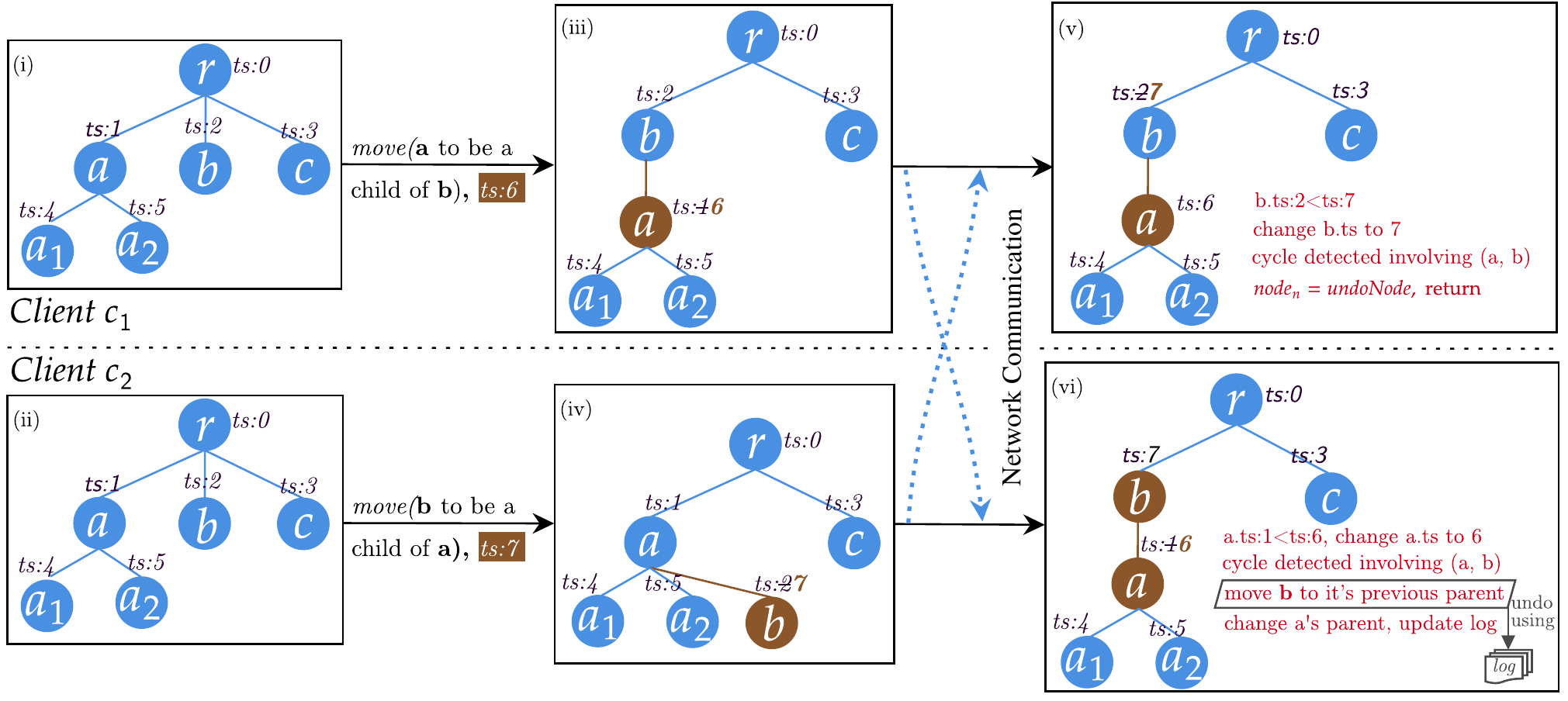}}%\vspace{-.2cm}
	\caption{Preventing cycles in proposed approach.}
	\label{fig:proposed}
\vspace{.25cm}
    \hrule
\vspace{.3cm}
    %\fbox
	{\includegraphics[width=.68\textwidth]{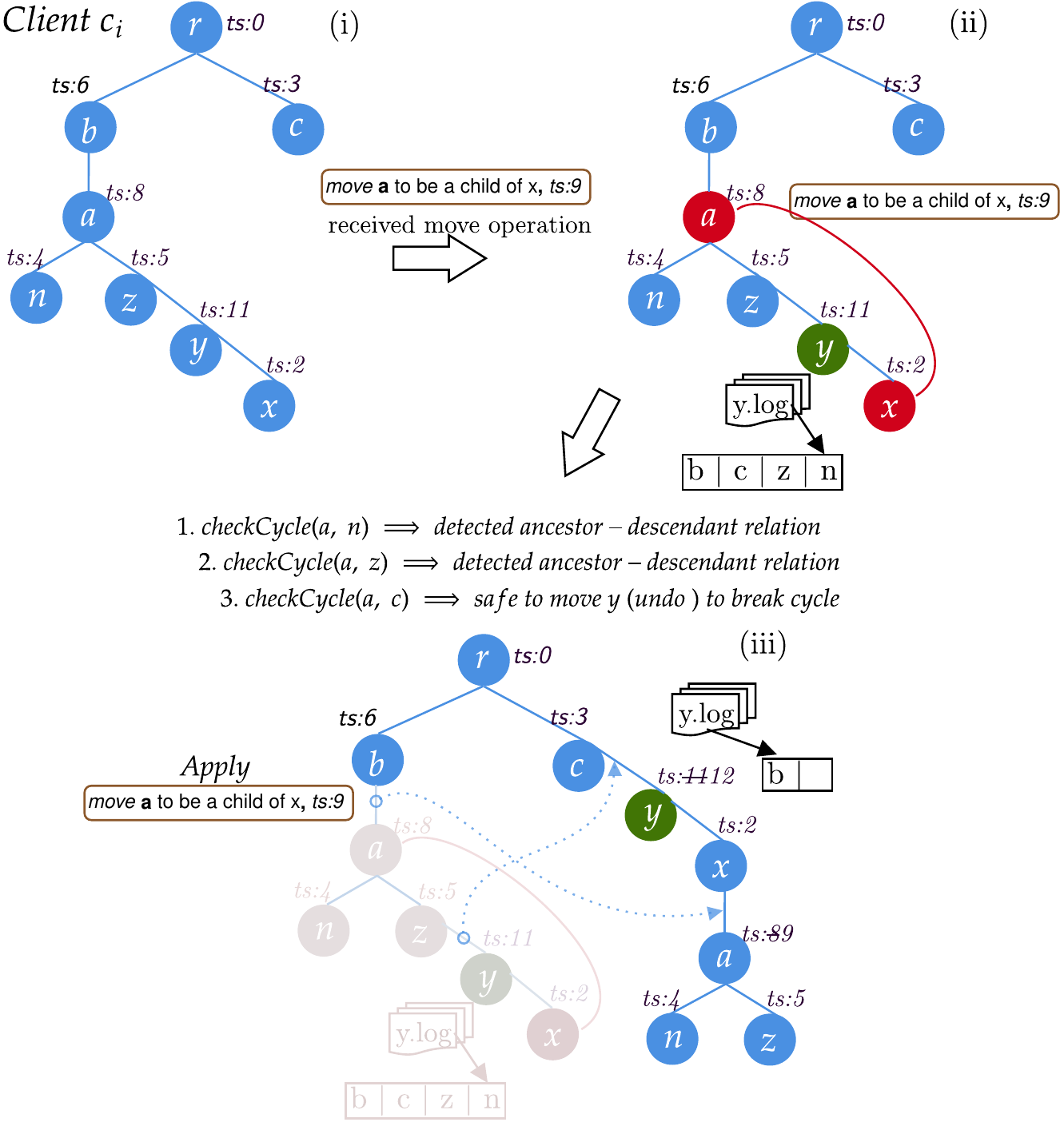}}%\vspace{-.42cm}
    \caption{Remote move \op{} handling at a replica.}
    \label{fig:exp2}
\end{figure}

%\vspace{.15cm}
\clearpage
\noindent
\textbf{Working Examples: }From \figref{cycle}, let us assume that initially, both clients (replicas) consist of the same tree with \timestamp{} as shown in \figref{proposed} (sub-figure (i) and (ii)). Each operation is assigned unique \timestamp{}. Client $c_1$ generates and applies the local operation \emph{move$_1\langle$ts:6, $n$:a, $p$:b$\rangle$}, i.e., \emph{move \circled{a} to be a child of \circled{b}} with \timestamp{} $move_1.ts:6$ as shown in (iii). Similarly, client $c_2$ generates and applies the local operation \emph{move$_2\langle$7, b, a$\rangle$}, i.e., \emph{move \circled{b} to be a child of \circled{a}} with \timestamp{} $move_2.ts:7$ as shown in (iv). %Note that operation at source replicas are safe since move operation does not produce any cycle. When clients receives operations from remote replicas, cycles may be formed.
As shown in \figref{proposed} (v) when client $c_1$ receives the \emph{move$_2\langle$7, b, a$\rangle$} operation from $c_2$, it executes the following steps:

\begin{enumerate}
    \item Operation \timestamp{} ($move_2.ts:7$) is not less then move node \timestamp{} ($b.ts:2$).% (\algoref{apply}, \lref{algo1-16}). 
    
    \item Change move node \timestamp{} to operation \timestamp{}, i.e., $b.ts:7$.% (\lref{algo1-18}).
    
    \item Cycle is detected involving ($a$, $b$).%at \lref{algo1-19}. 
    
    \item The node in the cycle with the highest \timestamp{} is \circled{b} that is found using \algoref{findLast}.% (\lref{algo1-20}).
    
    \item The node \circled{b} is moved to its previous parent which is \circled{r}.% (\lref{algo1-wloop} $-$ \lref{algo1-21}).
    
    \item The compensation operation is propagated to other replicas to ensure that every replica has seen and applied the same set of operations.% at \lref{algo1-23}.
    
    \item Since $node_n$ (i.e., node received in move operation) is same as node to be moved back%(i.e., \emph{undoNode}) at \lref{algo1-24}
    , so algorithm returns.% at \lref{algo1-25}. 
\end{enumerate}

Similar steps are followed when operation \emph{move$_1\langle$6, a, b$\rangle$} from $c_1$ is received at $c_2$ (see \figref{proposed} (vi)). Except for case (7), where $node_n$ is not the same \emph{undoNode}. %, i.e., \lref{algo1-24} returns false. 
%Hence, \lref{algo1-26} will be executed, and 
Operation is applied on node \circledm{b}. In summary, node \circledm{b} is moved back to its previous parent \circled{r} using information stored in the \logFile{}, then received operation is applied, i.e., \circledm{a} is moved as a child of \circledm{b}, and the \logFile{} is updated. %Here at the end, the tree at both the clients (replicas) converges to the same state.

%{\color{red}Will add more explanation of \figref{proposed} in detail here.}
\iffalse
\begin{figure}[!ht]%[16]{r}{0.4\textwidth}
    \centering
    {\includegraphics[width=.7\textwidth]{figs/fig-exp-3.pdf}}%\vspace{-.42cm}
    \caption{Remote move \op{} handling at a replica.}
    \label{fig:exp2}
\end{figure}
\fi
Note that when a cycle is detected, the algorithm identifies the node with the latest \timestamp{} and move that node back (e.g., \circled{b} in \figref{proposed} (vi)) to the previous parent, where it is safe. In the \algoref{apply}: \lref{algo1-wloop} $-$ \lref{algo1-wloopend} tries to identify the previous parent of the node (e.g., \circled{b} in \figref{proposed} (vi)) where it can be moved back safely, and cycle can be broken. We are storing `m' \newtext{(a constant number)} previous parent for  tree node in an adjacency list in the \logFile{}. %Identifying the optimal value of `m' is left as future work. 
%\vspace{.3cm}

Let us consider another example, as shown in \figref{exp2} (i), when a replica (i.e., $Client\ c_i$) receive an \op{} to \emph{move \circled{a} to be a child of \circled{x}}. As shown in \figref{exp2} (ii), since \circled{a} is an ancestor of \circled{x} it will be detected that this \op{} can form a cycle. So algorithm finds the node with the highest timestamp in the potential cycle between \circledred{a} and \circledred{x}. Here, \circledgreen{y} has the highest timestamp; therefore, \circledgreen{y} can be moved to one of its previous parents to break the cycle. The algorithm, checks if any of its previous parents are safe (i.e., the previous parent is not a descendant of \circled{a}). The check fails for \circled{n} and \circled{z} but \circled{c} is a safe previous parent. Hence move \circledgreen{y} to be a child of \circled{c}, as shown in \figref{exp2} (iii). As a result of this, \circled{x} is no longer a descendant of \circled{a}. So apply the original \op{} to move \circled{a} to be a child of \circled{x}.

%We store the previous parents of all nodes in an adjacency list in the \logFile{}. 
\newtext{Storing fixed `m' previous parents for each node will be adequate by considering storage; moreover, increasing the value of `m' increases the search time. For current experiments, `m' is fixed to $5$. Identifying the optimal number of previous parents (i.e., `m') is left as future work. If previous parents in \logFile{} are too small for the node to be moved to break the cycle, it is moved under a special node known as the \conf{} (a child of the root) to break the cycle. The \conf{} is special node that cannot be moved, it ensures that it will always be free from cycles.} %The replica needs to send an operation to move that tree node to \conf{}\clrtxt{/\emph{previous parent}} to other replicas since not every replica needs to form this cycle as operations are applied in a different order at each replica.
In case if the number of previous parents (or `m') for a node to be moved is deleted (by moving under \trash{}), we still can move the node under the previous parent \newtext{(deleted node in this case)}. Next, the clients have the choice to change location again as the nodes attached to \trash{} have not been deleted permanently. %We move the tree nodes with the highest \timestamp{} in the cycle to a child of the \conf{}. 
\newtext{Even when all previous parents are permanently deleted or \logFile{$_x$} for a node $x$ is empty, we can still move that node under \conf{} to break the cycle.} %However, for practical reasons, when storing `m' previous parents are not adequate, i.e., the value of `m' is very small, the conflicting node can be moved to a special node called \conf{}$-$a child of the \rt{}.

%We store the previous parents of all nodes in an adjacency list in the \logFile{}. \newtext{Storing fixed `m' previous parents for each node will be adequate by considering storage; moreover, increasing the value of `m' increases the search time. However, for practical reasons, when storing `m' previous parents are not adequate, i.e., the value of `m' is very small, the conflicting node can be moved to a special node called \conf{}$-$a child of the \rt{}.} %, the approach is as follows. %To avoid cycle following approach is followed: 

%If we find a cycle, we find the node in the cycle with the highest timestamp. In other words, the node that was moved last in the global order. We make another special node called \conf{} that is a child of the \rt{} node. We move the tree nodes with the highest \timestamp{} in the cycle to be a child of the \conf{} node. Since \conf{} cannot be moved, we can be sure that it will always be free from cycles. The replica needs to send an operation to move that tree node to \conf{} to other replicas since not every replica needs to form this cycle as operations are applied in a different order at each replica.

\newtext{The following important question is,} how do we know which tree nodes are part of the cycle? As previously mentioned, the cycle is formed when the node moves to an ancestor of its new parent. Hence the cycle will lie between the new parent and the node to be moved. Since operations are applied one by one, an operation can almost form only one cycle. When a user sees nodes attached to \conf{}, they understand it resulted in the formation of a cycle and had to be resolved. A valid question could be why not just prevent the last operation for each replica that results in the cycle instead of looking for the global last operation. It is possible that since operations are applied across replicas in different orders, they will result in each replica moving a different tree node to \conf{} \clrtxt{ / \emph{previous parent}}. This means that a large number of operations could be ignored. However, by looking for the global last operation, only the effect of one operation gets ignored.

\vspace{.15cm}
\noindent
\textbf{Globally Unique Timestamps: }We use Lamport clock~\cite{Lamport1978clock} on each replica for timestamping operations. However, this alone will not make it globally unique. Hence, we use the \emph{replica id} for tiebreakers when the Lamport \timestamp{s} are equal; together, they globally unique. As an alternative to the Lamport clock, the hybrid logical clock~\cite{kulkarni2014HLC} can be used that provides the unique \timestamp{s}.

\vspace{.15cm}
\noindent
\textbf{Difficulties: }\Trash{} can grow indefinitely. There can be a permanent \emph{delete} that recursively deletes from the leaf nodes to maintain consistency. Similar to the \emph{rm -r} command in Linux. However, this can lead to the case where a permanently deleted node is in the log of previous parents of other nodes. We will have to skip that parent and continue to the next previous parent in such cases. If none of the previous parents are safe or are permanently deleted, we move the node to be the child of \conf{}. As we are storing only the last $m$ previous parents and then moving to \conf{}, we are missing out on moving to the older positions if none are safe. %However, in a practical scenario, even just 2 or 3 previous parents should be enough in most cases. In the rest of them, 
A sound argument can be made that instead of moving a node to ancient locations, it is better to move to the \conf{} to notify the user that the \emph{move} operation on the following node was unsafe. Users can find another suitable location to move it to. Implementing permanent delete is left as future work.

Another important point is that, in the case of a cycle \newtext{due to remote operations}, the proposed approach requires propagating the undo operation to other replicas (requires one undo operation per remote move operation) that may be an additional message cost. However, it decreases many undo and redo operations to just one compensation operation at each replica, compensating for additional message costs. To summarize, we provide a replicated tree that can support efficient and highly available move operation.% without any coordination, total ordering.

%\subsection
%\vspace{.2cm}
%\noindent
\section{Proof of Correctness}
\label{sec:pc}
%\textbf{Correctness: }
This section provides the formal proof that all the replicas eventually converge to the same tree (state) and maintain the tree structure through optimistic replication. All our operations (\emph{insert, delete, move}) are just changing the parent to which the node is attached. 

\vspace{.2cm}
\begin{lemma}[Duplication]
\label{lem:duplicate}
\emph{A tree node will never be located at multiple different positions.}
\end{lemma}
\vspace{.15cm}
\begin{proof}
    For every node, we only store a single value for its parent. It must have more than one parent to be duplicated at multiple positions. However, that is not possible in our approach since we only store a single parent based on the last written win semantic.
    %We store a single value for the parent of each node. In order to be duplicated at multiple positions, it must have more than one parent. This is not possible in our approach, as we only store a single parent based on the most recently written win semantics.
	%For every node, we only store a single value for its parent. So for it to be duplicated at multiple positions, it needs to have more than one parent. However, that is not possible in our approach since we only store a single parent based on the last write wins semantics. %\qed 
\end{proof}
%\vspace{.2cm}
%\noindent
%\textbf{Definition 1 \emph{Duplication}: }\emph{A tree node will never be located at multiple different positions.}

\vspace{.2cm}
\begin{lemma}[Cycle]
\label{lem:cycle}
\emph{No operation will result in the formation of a cycle.}
\end{lemma}
\vspace{.15cm}
\begin{proof}
    Before applying any operation, our algorithm tries to detect if it will form a cycle. Assume we have an operation of the form $\langle ts, n, p \rangle$ where $ts$ is a timestamp, $n$ is the tree node to be moved, $p$ is the parent. As previously stated, the operation will only form a cycle if $n$ is an ancestor of $p$. 
    
    We traverse all the way up from $p$ to \rt{}, and in case we do not find $n$ in that path, it implies $n$ is not an ancestor of $p$. Hence the operation is safe to apply and will not form a cycle. If we find $n$ is an ancestor of $p$, we will apply an alternate compensation operation. Hence we never apply an operation where $n$ is the ancestor of $p$.
    
    So now, for a cycle to exist, it needs to be formed from an operation where $n$ is not the ancestor of $p$. How to prove that if $n$ is not an ancestor, it will not form a cycle?
    
    If $n$ and $p$ need to form a cycle, there should be a path from $n$ to $p$ and $p$ to n. However, applying the operation will only create a path from $n$ to $p$. The path from $p$ to $n$ needs to exist before, and such a path will only happen if $n$ is an ancestor of $p$. %\qed 
\end{proof}

\vspace{.2cm}
\begin{lemma}[Forest]
\label{lem:forest}
\emph{The tree will never be split into multiple forests.}
\end{lemma}
\vspace{.15cm}
\begin{proof}
    We always maintain a parent for every tree node other than \rt{}, and we do not allow operations that have the same parent and tree node value. The tree could be split into forests if there is some node without a parent or a cycle. We have already shown that cycles cannot be formed, and our algorithms always maintain a parent for each tree node other than \rt{}. %\qed 
\end{proof}
%\vspace{-.25cm}

\vspace{.2cm}
\begin{theorem}[Safe]
\label{thm:safe}	
\emph{A previous parent is said to be safe if it is not in the sub-tree of the node to be moved in the move operation.}
\end{theorem}
\vspace{.15cm}
\begin{proof}
    \clrtxt{
        Since the node with the highest timestamp in the cycle is moved back to the previous parent, i.e., the previous parent must not be in the sub-tree of the node to be moved in a move operation, moreover, when there is no previous parent such that it is not in the sub-tree, then \conf{} (special child of the \rt{} that can not be moved) is assigned as the previous parent. So one of the nodes in the cycle is always moved back to a node which is not in the sub-tree of the node to moved. As a result of this the new parent of the node to be moved will no longer be in its sub-tree. So there is no ancestor-descendant relation between the node to be moved and the new parent and the operation is safe to be applied.
        %Since the node with the highest timestamp in the cycle is moved back to the previous parent, i.e., the previous parent must not be in the sub-tree of the node to be moved in a move operation, moreover, when there is no previous parent such that it is not in the sub-tree, then \conf{} (special child of the \rt{} that can not be moved) is assigned as the previous parent. So one of the nodes in the cycle is always moved back, and the cycle will be broken, which guarantees that there will be no new cycle due to move back operation, from Lemmas \ref{lem:cycle1}. Hence the move back node is eternally moved to a safe location to break the cycle.
        %\qed 
    }
\end{proof}

\vspace{.15cm}
Having explained the lemmas and theorem, we now explain the main theorem. 

\vspace{.2cm}
\begin{theorem}[Convergence]
\label{thm:converge}	
\emph{All the replicas that have seen and applied the same set of operations will converge to the same tree.}
\end{theorem}
\vspace{.15cm}
\begin{proof} 	
	Say two replicas $r_1$ and $r_2$ have seen the same set of operations. They will have the same parent for each node as the operation with the latest \timestamp{} taken as the parent.
	
	%From Lemmas \ref{lem:duplicate}, \ref{lem:cycle}, \ref{lem:forest} we get that the tree structure is maintained. 
	From Lemmas~\ref{lem:duplicate}, \ref{lem:cycle}, \ref{lem:forest}, and Theorem~\ref{thm:safe}, we get that the tree structure is maintained and there will be no cycles in the tree. Next, assume that replicas $r_1$ and $r_2$ have seen the same set of updates but have different parents for a key ($k$). Suppose the replica $r_1$ for $k$ has \timestamp{} $ts_1$ and parent $p_1$. The replica $r_2$ for $k$ has \timestamp{} $ts_2$ and parent $p_2$. We know $ts_1 \neq ts_2$ since we are using globally unique \timestamp{}s (ties are broken by replica id), and if they were equal, then parents would have been the same. This implies that either $ts_1 < ts_2$ or $ts_1 > ts_2$. This means that one of the replicas has not applied the latest \timestamp{}. However, according to correctness of our algorithm, it was supposed to do that. Hence this is not possible. It means the initial assumption was wrong that the replicas have different parents for the same key or have seen the same set of updates. %\qed 
\end{proof}
    \section{Performance Evaluation}
\label{sec:results}
This section presents the implementation details (\subsecref{imp}) and performance comparison (\subsecref{analysis}) of the proposed approach with the state-of-the-art approach by Kleppmann et al.~\cite{Kleppmann2020AHM} in a geo-replicated setting at three different continents to demonstrate the usefulness of the proposed approach. 

\subsection{Implementation}
\label{subsec:imp}
We have implemented the proposed algorithm in Golang~\cite{golang} and wrapped it in gRPC~\cite{grpc} network service to deploy at three different geo-location (Western Europe, Southeast Asia, and East US) on Microsoft Azure Standard E2s\_v3 VM instances, each consisting of 2 vCPU(s), 16 GiB of memory, and 32 GiB of temporary storage running Ubuntu 20.04 operating system and Intel Xeon Platinum 8272CL processor.\footnote{Source code: \url{https://github.com/anonymous1474}}

\begin{table}[!t]
    %\vspace{.1cm}
    \caption{Network latency (ms) between different replicas}%\vspace{-.3cm}
    \label{tab:latency}
    \centering
    \resizebox{\columnwidth}{!}{%
        \begin{tabular}{c|c|c|c}
            \hline
           & \textbf{~~~~~~~~~US East~~~~~~~~~} & \textbf{~~~~~~~~~West Europe~~~~~~~~~} & \textbf{~~~~~~~~~Southeast Asia~~~~~~~~~}\\\hline
            \textbf{US East} & 0 & 41 & 111\\
            \textbf{West Europe} & 41 & 0 & 79\\
            \textbf{~~~~~Southeast Asia~~~~~} & 111 & 79 & 0\\\hline
        \end{tabular}
    }
\end{table}

The network latency from US East to Southeast Asia is 111 ms, the maximum, and West Europe to US East is 41 ms which is the minimum. \tabref{latency} shows the network latency's between different geo-locations chosen for the experiments.  %displayed in \tabref{latency}. 
However, network latency's are not considered in the final results because the %\Sout{\emph{response time}$-$}
\emph{time to apply} a move operation (local or remote) is computed at each replica. We ran the experiments 7 times. The first two runs are considered warm-up runs, hence each point in the plot is averaged over five and across the different replicas.
%We assume a reliable network. The network latency between different VM instances is shown in the \tabref{latency}, however not considered in the final results, since the \emph{response time}$-$time to apply move operation (local or remote) is computed at each replica. We ran the experiments 12 times, out of which the first two runs are considered warm-up runs, hence each point in the plot is averaged over ten.
The synthetic workload is used for the experiments consisting of insert, delete, and move operations. Initially, the tree is empty and based on the experiment, the number of nodes varies in the tree. The node is identified by \emph{key} or \emph{node id} an integer. The Lamport clock~\cite{Lamport1978clock} is used to timestamp each operation. 

Each replica generates and applies local operations, subsequently asynchronously propagating and receiving the operations to/from the other two replicas. Replica generates the move operation by selecting tree nodes uniformly at random from the tree size. Each replica generates $(\frac{1}{3})^{rd}$ of the total number of operations applied and receives $(\frac{2}{3})^{rd}$ of the operations from the other two replicas. %Each replica generates and applies local operations, subsequently asynchronously propagates and receives the operation to/from the other two replicas. Replica generates the move operation by selecting tree nodes uniformly at random from the tree size. From the total number of operations applied, each replica generates $(\frac{1}{3})^{rd}$ of the operations and receives $(\frac{2}{3})^{rd}$ of operations from the other two replicas. %Replica generates and applies the local operation as soon as a new operation is generated and propagates the local operations to other replicas. 
When a replica receives a remote operation, it applies, and in case of any undo operation due to cycle, identifies the appropriate previous parent (for the node with the higher timestamp in the cycle)% to perform undo \op
. Once the appropriate parent is identified, the node is moved as a child and the undo operation is sent to other replicas as per the protocol. %The local operation does not lead to any cycle since operation timestamps at local replicas are generated in increasing order and take less time than the remote operation, however, the cycle check is performed for local operations as well. 
Note that our experimental workload is more conservative and contains more conflict than the real-time workload. Further, move operations conflict only with other move or delete operations; they do not conflict with other operations (such as updating a value at a node in the tree or inserting a new subtree).

\subsection{Results and Analysis}
\label{subsec:analysis}
\begin{figure}[!tb]
	\centering
	\begin{tikzpicture}[scale = .5, every node/.append style = {scale = .8}, font = \Huge]
	\begin{axis}[
			base,
			%title	= {Time to apply remote move operation},
			ymax	= 500,
			ymin	= 0,
			%xmax	= 620,
			xlabel	= Tree size (\# nodes),
			xtick	= {200,500,1000,2000},
			ytick	= {0,100,200,300,400,500},
			ylabel	= Average \# Conflicts\\(undo-redo operations),
			ylabel style = {align=center}, 
			legend style = {at = {(0.58,1.01)}, anchor = north, legend columns = 1},
			every axis plot/.append style = { line width = 2.5pt},
			/pgfplots/legend image code/.code = {%
					\draw[mark repeat=2,mark phase=2,#1] 
					plot coordinates {
						(0cm,0cm) 
						(0.5cm,0cm)
						(1.0cm,0cm)
						(1.5cm,0cm)
						(2.0cm,0cm)%
					};
				},	
		]
		
		\addplot[ color=black, mark=star, mark options={fill=black,solid }, mark size=7pt, dash pattern=on 1pt off 3pt on 3pt off 3pt, smooth]
		table[ x index=0, y index=1, col sep=space ] {exp/exp4.txt};\label{plot:line1}
		\node [align=center,anchor=west,scale=.7] at (225,430){427};
		\node [align=center,anchor=west,scale=.7] at (1950,85){47};
		
		\addplot[ color=red, mark=square, mark options={fill=red!60,solid}, mark size=4.5pt, dotted, smooth]
		table[ x index=0, y index=2, col sep=space ] {exp/exp4.txt};\label{plot:line2}
		\node [align=center,anchor=west,color=red,scale=.7] at (5,336){332};
		\node [align=center,anchor=west,color=red,scale=.7] at (2040,48){48};
		
		\addplot[ col2, mark=triangle, mark options={fill=col2!70,solid}, mark size=5pt, loosely dotted,smooth]
		table[ x index=0, y index=3, col sep=space ] {exp/exp4.txt};\label{plot:line3}
		\node [align=center,anchor=west,color=col2,scale=.7] at (5,397){394};
		\node [align=center,anchor=west,color=col2,scale=.7] at (1950,20){44};
		
		\node [align=center,anchor=west,scale=.8,fill=white] at (670,320){\textbf{\# operations: 5000}};
		\node [align=center,anchor=west,scale=.8,fill=white] at (670,280){\textbf{Fixed 500 operations per sec.}};
		\legend{~Western Europe~~,~Southeast Asia~~~, ~East US~~~~~~~~~~}
	\end{axis}
\end{tikzpicture}
%\vspace{-.3cm}
\caption{Average number of conflicts (undo and redo operations) at different replicas in proposed approach.}
\label{fig:exp4}
%\vspace{-.7cm}
\end{figure}
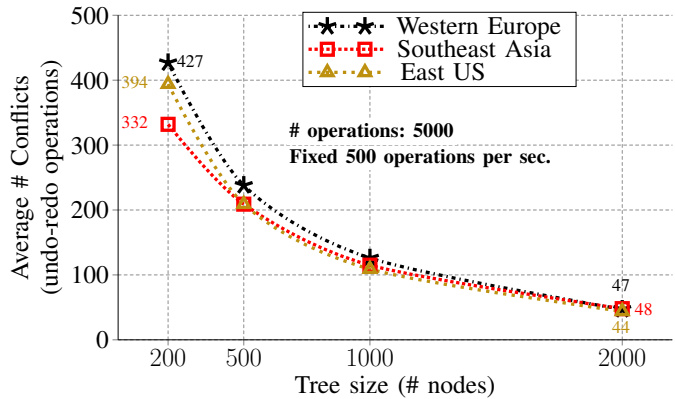

We performed two kinds of experiments as shown in \figref{exp1-2} and \ref{fig:exp4}. In \figref{exp1-2}, we show experiments to evaluate the performance when the number of tree nodes is fixed to 500 while the number of operations that a single replica is issuing per second is varying. The x-axis is the \emph{operations per second} varied from 250 to 5000 while the y-axis is the time taken to apply an operation (local or remote) at a replica. 

In \figref{exp4}, the experiment is a function of nodes in the tree to conflicting operations when operations are fixed to 15K (5K operations per replica) and fixed to 500 operations per second on three different geo-locations. This experiment shows the number of undos and redos operations in the proposed approach at different replicas. 

Having explained the high-level overview of the experiments, we now go into the details. \figref{exp1-2} depicts the average time to apply a local and remote move operation. In this experiment, the number of operations per second is varied from 250 to 5K, while the number of nodes in the tree is kept constant at 500. As illustrated in \figref{exp1-2}(a), the average time to apply a local move operation at a replica is consistent and not so significant between both approaches when the number of operations increases. The apply time for a local move operation drops as the number of operations increases. Kleppmann et al.~\cite{Kleppmann2020AHM} also observed similar trends for local operations. However, the time to apply a remote move operation is almost constant in the proposed approach, while Kleppmann's approach has the opposite trend; the time increases with operations per second for remote move operations as shown in \figref{exp1-2}(b). There is a significant performance gap for the remote operation in both the approaches; this is because the number of compensation operations (undo/redo) by Kleppmann's approach is $\approx$ 200 undo and redo operations for every remote operation a replica receives while in our case it is only 1 that too whenever there is a cycle (conflict).
%{\color{red}On average, the proposed approach outperforms Kleppmann's approach by $\approx XX\times$ to apply remote move operation at a replica. Next, we explain the reason behind this performance increase.}

\begin{figure}[!tb]
	\begin{tikzpicture}[scale = .5, every node/.append style = {scale = .8}, font = \Huge]
		\begin{axis}[
			base,
			%nnc,
			title	= {(a) Local move operation apply time},
			%ymax	= 3,
			ymin	= -0.5,
			%xmax	= 620,
			xlabel	= operations per second,
			xtick	= {250,1000,2000,5000},
			ytick	= {0.0,2.0,4.0,6.0,8.0,10.0},
			ylabel style={align=center}, 
			ylabel 	= Average Time to Apply\\a Local Operation ($\mu$s),%Operation Apply Time ($\mu$s), 
			%Average Time to\\Apply an Operation ($\mu$s),
			%y label style = {at = {(-0.14,.95)}, anchor = south east},
			legend style  = {at = {(0.6,1)}, anchor = north, legend columns = 1},
    		every axis plot/.append style = { line width = 3.2pt},
			/pgfplots/legend image code/.code = {%
					\draw[mark repeat=2,mark phase=2,#1] 
					plot coordinates {
						(0cm,0cm) 
						(0.5cm,0cm)
						(1.0cm,0cm)
						(1.5cm,0cm)
						(2.0cm,0cm)%
					};
				},
				nodes near coords,
    		    %every node near coord/.append style = { xshift = 14pt, yshift = 6pt, rotate = 45, font = \large }, 
			]
    		\addplot[color=blue, mark=+, mark options={fill=blue,solid }, mark size=5pt, dash pattern=on 1pt off 3pt on 3pt off 3pt,smooth,every node near coord/.append style = { xshift = -8pt, yshift = -30pt, rotate = 45, font = \huge }]
    		table[ x index=0, y index=1, col sep=space ] {exp/exp2-local.txt};
    		
    		\addplot[color=black, mark=diamond*, mark options={fill=white,solid}, mark size=5pt, dotted,smooth,every node near coord/.append style = { xshift = 24pt, yshift = 18pt, rotate = 45, font = \huge }]
    		table[ x index=0, y index=2, col sep=space ] {exp/exp2-local.txt};

			%\addplot  [color = nodeblue, fill = nodeblue!50] 
			%table [x index=0, y index=1,col sep=space] {exp/exp2-local.txt};
			
			%\addplot [color = ngr, fill = ngr!50] 
			%table [x index=0, y index=2,col sep=space] {exp/exp2-local.txt};
		
			\legend{~Proposed~~~~~~~~~~~~~~~~,~Kleppmann et al.~\cite{Kleppmann2020AHM}}
		\end{axis}	
	\end{tikzpicture}
\hspace{.2cm}
	\begin{tikzpicture}[scale = .5, every node/.append style = {scale = .8}, font = \Huge,]
		\begin{axis}[
	        grid,
			base,
			%nnc,
			title	= {(b) Remote move operation apply time},
			%ymax	= 30,
			ymin	= -170,
			%xmax	= 620,
			xlabel	= operations per second,
			xtick	= {250,1000,2000,5000},
			ytick	= {0,200,400,600,800,950},
			%xtick	= {100,200,300,400,500},
			%ylabel	= Response Time ($\mu$s),
			ylabel style={align=center},
			ylabel 	= Average Time to Apply\\a Remote Operation ($\mu$s),
			legend style  = {at = {(0.4,1)}, anchor = north, legend columns = 1},
    		every axis plot/.append style = { line width = 3.2pt},
			/pgfplots/legend image code/.code = {%
					\draw[mark repeat=2,mark phase=2,#1] 
					plot coordinates {
						(0cm,0cm) 
						(0.5cm,0cm)
						(1.0cm,0cm)
						(1.5cm,0cm)
						(2.0cm,0cm)%
					};
				},
				nodes near coords,
    		    %every node near coord/.append style = { xshift = 14pt, yshift = 14pt, rotate = 70, font = \large }, 
			]
    		\addplot[color=Mahogany, mark=o, mark options={fill=Mahogany,solid }, mark size=4pt, dash pattern=on 1pt off 3pt on 3pt off 3pt,smooth,every node near coord/.append style = { xshift = -3pt, yshift = -25pt, rotate = 60, font = \huge }]
    		table[ x index=0, y index=1, col sep=space ] {exp/exp2-remote.txt};
    		
    		\addplot[color=DarkOrchid, mark=triangle, mark options={fill=DarkOrchid!60,solid}, mark size=5pt, dotted,smooth,every node near coord/.append style = { xshift = 15pt, yshift = 30pt, rotate = 60, font = \huge }]
    		table[ x index=0, y index=2, col sep=space ] {exp/exp2-remote.txt};

		%	\addplot  [color =black,fill=black!50] 
		%	table [x index=0, y index=1,col sep=space] {exp/exp2-remote.txt};
			
		%	\addplot 
		%	table [x index=0, y index=2,col sep=space] {exp/exp2-remote.txt};
			
			\legend{~Proposed~~~~~~~~~~~~~~~~,~Kleppmann et al.~\cite{Kleppmann2020AHM}}
		\end{axis}
	\end{tikzpicture}

	%\vspace{-.2cm}
	\caption{Average time to apply a move operation.}\vspace{.5cm}
	\label{fig:exp1-2}
\end{figure}

As shown in \figref{exp1-2}(b), Kleppmann's approach attains a maximum apply time of 933.53$\mu$s over a remote move operation at 5K operations per second. In comparison, the minimum is 81.69$\mu$s at 250 operations per second; the minimum time is $\approx$ 14.63$\times$ higher than the maximum time of 5.58$\mu$s at 250 operations by the proposed approach. It can be seen that the proposed approach achieves, on average, a speedup of 1.34$\times$ for the local move operation, while 68.19$\times$ speedup for the remote move operation over Kleppmann's approach. Hence, the proposed approach is much faster in applying remote operations, and the difference in time only increases with an increase in the rate of operations per second. This shows the performance benefits of the proposed approach. In Kleppmann's approach, as the rate of operation generation increases, the number of operations in flight also increases. As a result, the compensation cost will be high, i.e., a more significant number of undo and redo operations. In contrast, even if the rate increases in the proposed approach, the compensation cost remains the same in the proposed approach. Next, we explain the reason behind this performance gain.

In \figref{exp4}, a line chart depicts the average number of conflicts (undo and redo operations) at different replicas in the proposed approach. In this experiment, we fixed the number of local operations to 5K per replica (total 15K operations) and the operation interval to 10 milliseconds while varying the number of nodes in the tree from 200 to 2K. This experiment is performed to demonstrate the number of undos and redos operations performed by each replica and the system performance when the tree size is changed. The maximum number of conflicts is $\approx$ 427 when the number of nodes in the tree is 200, but it drops to $\approx$ 47 when the number of nodes in the tree is 2K, as shown in the \figref{exp4}. These numbers are much smaller than the number of undos and redos by Kleppmann's approach, which roughly equals 2M (million) for 10K remote move operations at a replica in the worst case. Since the average number of undos and redos per remote move operation in Kleppmann's approach is $\approx 200$~\cite{Kleppmann2020AHM}; as a result, we can see that the proposed approach significantly improves the remote move operations apply time than Kleppmann's approach.

%The average number of conflicts (undo and redo operations) at different replicas in the proposed approach is shown as a line chart in \figref{exp4}. In this experiment, we fixed the number of operations to 4000 and the operation interval to 10 ms while varies the number of nodes in the tree from 200 to 2000. We performed this experiment to demonstrate the number of undos and redos operations performed by each replica and the system performance with the change in the tree size. We can observe from the \figref{exp4} that the maximum number of conflicts is $\approx$125 when the number of nodes is 200 while it drops to $\approx$10 when nodes in the tree are 2000. These numbers are significantly less than the average number of undo and redo operations by Kleppmann's approach, which roughly will be 533200 for $\approx$2666 remote operations at a replica in the worst case. Because of this reason, we can see the clear benefit in the response time of remote move operation by the proposed approach. 
    \section{Conclusion and Future Directions}
\label{sec:conc}
%CRDTs such as counters, lists, graphs, trees, etc., become an essential component of many distributed collaborative applications. Moving a node is an essential operation in tree-based collaborative applications.

We proposed a novel algorithm for efficient move operations on a replicated tree structure. The proposed technique ensures that replicas that have viewed and applied the same set of operations will eventually converge to the same state. Moreover, it does not require active cross-replica communication, making it highly accessible even during network partitions. We have followed a last write win scheme on globally unique timestamps. The proposed technique requires a single compensating operation to undo the effect of the cyclic operation. It achieves an average speedup of $\approx$ 68.19$\times$ over the state-of-the-art approach. We have stored a constant number of the previous parents for every node. Identifying the optimal number of previous parents is left as future work. Implementing an efficient move operation on other replicated data structures could be an exciting area to explore. Also, performing operations on a range of elements in the list and tree CRDTs, or applying operations in a group from the same replica, is another potential direction to pursue. %and left as future work.
    \bibliographystyle{IEEEtran}
    \bibliography{references}

\end{document}